\documentclass[12pt]{iopart}

\usepackage{iopams}
\usepackage{epsf}
\usepackage{graphicx}
\usepackage{fancyhdr}

\newcommand{\ndpar}[3] { \frac{\partial^{#3} {#1} } {\partial #2 ^{#3} }}

\newcommand{\eqref}[1] {(\ref{#1})}

\newcommand{\II}[0] {\mbox{i}}

\newcommand{\text}[1] {\mbox{#1}}

\newcommand{\dis}[0] {\displaystyle}

\newcommand{\abs}[1]{\left| #1 \right|}

\begin{document}

\title[A transfer matrix method for the analysis of fractal quantum
potentials] {A transfer matrix method for the analysis of fractal
quantum potentials}

\author{Juan A. Monsoriu$^1$\footnote[7]{To whom
correspondence should be addressed (jmonsori@fis.upv.es)},
Francisco R. Villatoro$^2$, Mar{\'\i}a J. Mar{\'\i}n$^3$, Javier
F. Urchuegu{\'\i}a$^1$ and Pedro Fern\'andez de C\'ordoba$^4$}

\address{$^1$ Departamento de F{\'\i}sica Aplicada,
 Universidad Polit\'ecnica de Valencia, E-46022 Valencia, Spain}

\address{$^2$ Departamento de Lenguajes y Ciencias de la Computaci\'on,
  Universidad de M\'alaga, E-29071 M\'alaga, Spain}

\address{$^3$ Departamento de Termodin\'amica,
 Universitat de Val\`{e}ncia, E-46100 Burjassot, Spain}

\address{$^4$ Departamento de Matem\'atica Aplicada,
 Universidad Polit\'ecnica de Valencia, E-46022 Valencia, Spain}

\begin{abstract}
The scattering properties of quantum particles on fractal
potentials at different stages of fractal growth are obtained by
means of the transfer matrix method. This approach can be easily
adopted for project assignments in introductory quantum mechanics
for undergraduates. The reflection coefficients for both the
fractal potential and the finite periodic potential are calculated
and compared. It is shown that the reflection coefficient for the
fractal has a self-similar structure associated with the fractal
distribution of the potential.
\end{abstract}


\submitto{\EJP}

\maketitle

\section{Introduction}
\label{Intro}

Both quantum mechanics and elementary solid state physics courses
illustrate the energy band structure in solids through the
one-dimensional Kronig-Penney model that consists of a periodic
configuration of square-well
potentials~\cite{KronigPenney,Liboff}. This problem is usually
solved by matching the boundary conditions of the wavefunctions at
the cell boundaries, thus requiring the computation of the
determinant of a $4\times 4$ matrix~\cite{Szmulowicz}. Recently,
some less tedious approaches have been proposed which usually can
be readily adapted to finite periodic potentials~\cite{Sprung}.
Among these methods, those based on the transfer matrix approach
which only uses $2\times 2$ matrix operations in a purely
algebraic way are the most appropriate ones for
beginners~\cite{Griffiths}. Moreover, this method allows to
introduce a numerical method based on a piecewise constant
approximation~\cite{Kalotas} for a general potential, the analysis
of defects on slightly aperiodic potentials, and even the
consideration of more complicated potentials. Among the last,
fractal potentials is the one which we considered here.

In recent years the study of fractals has attracted much attention
because many physical phenomena, natural structures and
statistical processes can be analyzed and described by using a
fractal approach~\cite{Mandelbrot,Ficker}. From a mathematical
point of view, fractals are self-similar structures obtained by
performing a basic operation, called {\it generator}, on a given
geometrical object called {\it initiator}, and repeating this
process on multiple levels; in each one of them, an object
composed of sub-units of itself is created that resembles the
structure of the whole object. Mathematically, this property
should hold in all scales. However, in the real world, there are
lower and upper bounds over which such self-similar behavior
applies. Fractals are becoming a useful tool to be able to model
diverse physical systems~\cite{Berry,Karman}, and have new
technological applications~\cite{Saavedra,Monsoriu}.

In non-relativistic quantum mechanics, fractals have been used to
generate new solutions of the Schr\"odinger equation which are
continuous but nowhere differentiable wave
functions~\cite{Wojcik}, and models for the so-called fractal
potentials~\cite{Albeverio}. Fractal potentials allow the analysis
of quasi-periodic and nearly stochastic potentials using the
symmetries induced by the self-similar structure of the potential.
Here we consider the simplest fractal, the (triadic) Cantor set,
as a fractal potential for quantum scattering~\cite{Makarov} and
tunnelling~\cite{Chuprikov1,Chuprikov2}.

In this paper, we present a simple transfer matrix method to
obtain the scattering properties of Cantor set fractal potentials,
which can easily be automated by computers. The present method
makes easier for the comparison with the finite periodic case and
shows how the reflection coefficient for the fractal case has a
self-similar structure associated with the fractal distribution of
the potential. Moreoever, it can be easily implemented in any
computer language, e.g., the \emph{Mathematica} software package,
accesible to undergraduate students with only a basic programming
experience, so that it can be adopted for project assignments in
computer physics courses. Starting with the implementation of the
transfer matrix method for a potential barrier, the extension to
finite periodic potential is a straightforward one; its extension
to Cantor set potentials can be based on a recursive
implementation, involving the possible improvement of student
programming skills. Furthermore, these projects can introduce the
students to the analysis of computational complexity of
algorithms, since the Cantor set prefractal has a large number of
potential barriers and its simulation requires an exponential
number of matrix products.

This paper is organized as follows: The next section describes the
main facts about the transfer matrix method for quantum scattering
implemented by piecewise constant potentials. In Section~3, the
reflection coefficient for the tunnelling on both the finite
periodic potential as well as the Cantor set pre-fractals are
determined and compared. Finally, the last section is devoted to
the conclusions.

\section{The transfer matrix method in quantum scattering}
\label{Problem}

\begin{figure}
 \begin{center}
  \epsfbox{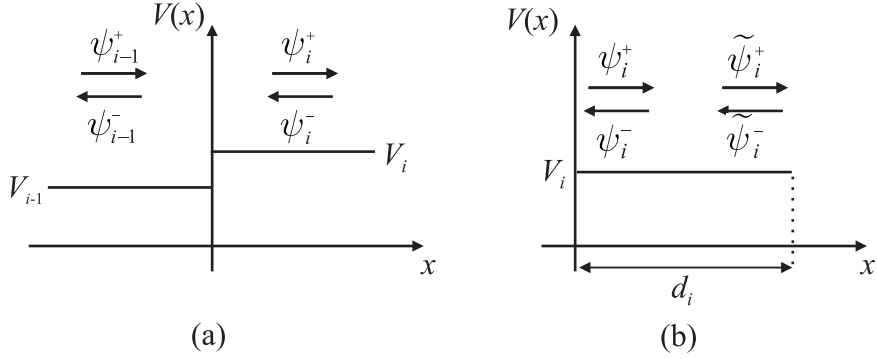}
 \end{center}
 \caption{\label{figuraMatrix}(a) Local scattering with the
 $i$-th interface of the piecewise constant potential
 among the values $V_{i-1}$ and~$V_{i}$. (b) Wave
 propagation through the constant potential $V_i$ as
  used to calculate the propagation matrix.}
\end{figure}

Let us consider the one-dimensional, steady-state, linear
Schr\"odinger equation
\begin{equation}
 \label{LSE}
 -\frac{\hbar^2}{2\,m}\,\ndpar{\psi(x)}{x}{2} + V(x)\,\psi(x) =
 E\,\psi(x),
\end{equation}
where $\psi(x)$, $m$ and $E$ are the wavefunction, mass and energy
of the particle, respectively. The constant $\hbar$ is Planck's
constant, and $V(x)$ is the quasiperiodic potential which can be
represented by a piecewise constant function.
Figure~\ref{figuraMatrix}(a) shows the quantum scattering at the
$i$-th interface between two successive constant values of the
piecewise potential, whose position, without loss of generality,
has been taken as $x=0$. In this figure, both $\psi^+_i$ and
$\psi^-_i$ are forward and backward plane wavefunctions,
respectively, on the region where the potential value is $V_i$,
and $\psi_i=\psi^+_i+\psi^-_i$. These wave functions are given by

\begin{equation}
 \psi^\pm_i = A^\pm_i\,e^{\dis  \pm\II\,k_i\,x},
\end{equation}
where $ k_i = \frac{1}{\hbar}\,\sqrt{2\,m\,(E-V_i)}$ is the local
particle momentum, and $A^\pm_i$ are integration constants to be
determined by applying the standard boundary conditions at the
interface. The continuity of the wavefunctions and the derivatives
at the boundary are given by
\begin{equation}
 \label{eq:Three}
 \eqalign{
 \psi_{i-1}(x=0)=\psi_{i}(x=0), \qquad
   A^+_{i-1} + A^-_{i-1} = A^+_i + A^-_i,\\
 \psi'_{i-1}(x=0)=\psi'_{i}(x=0), \qquad
   k_{i-1}\,A^+_{i-1} - k_{i-1}\,A^-_{i-1} = k_{i}\,A^+_i -
   k_{i}\,A^-_i,}
\end{equation}
where the prime denotes differentiation. Eq.~\eqref{eq:Three} is a
linear system of equations which can be written in matrix notation
as
\begin{equation}
 \left(\begin{array}{cc}
    1 &  1 \\ k_{i-1} & -k_{i-1}
 \end{array}\right)\,
 \left(\begin{array}{c}
    A^+_{i-1} \\ A^-_{i-1}
 \end{array}\right)
      =
 \left(\begin{array}{cc}
    1 &  1 \\ k_{i} & -k_{i}
 \end{array}\right)\,
 \left(\begin{array}{c}
    A^+_{i} \\ A^-_{i}
 \end{array}\right),
\end{equation}
and yielding
\begin{equation}
 \left(\begin{array}{c}
    A^+_{i-1} \\ A^-_{i-1}
 \end{array}\right)
    =
 D_{i-1}^{-1}\,D_{i}\,
 \left(\begin{array}{c}
    A^+_{i} \\ A^-_{i}
 \end{array}\right),
 \qquad
 D_i =
 \left(\begin{array}{cc}
    1 &  1 \\ k_{i} & -k_{i}
 \end{array}\right).
\end{equation}
Here on, the matrix $D_{i-1}^{-1}\,D_{i}$ is referred to as the
wave scattering matrix.

After crossing the $i$-th interface, the plane wave propagates
through the constant potential $V_i$ until it finds the next
interface at a distance $d_i$. Using the notation shown in
Figure~\ref{figuraMatrix}(b), this wavefunction is given by
\begin{equation}
 \widetilde{\psi}^\pm_i = A^\pm_i\,e^{\dis  \pm\II\,k_i\,d_i}\,e^{\dis  \pm\II\,k_i\,x}
 = \widetilde{A}^\pm_i\,e^{\dis  \pm\II\,k_i\,x},
\end{equation}
and a wave propagation matrix $P_i$ can be defined as
\begin{equation}
 \left(\begin{array}{c}
    \widetilde{A}^+_{i-1} \\ \widetilde{A}^-_{i-1}
 \end{array}\right)
    =
 \left(\begin{array}{cc}
    e^{\dis  \II\,k_i\,d_i} & 0 \\
    0  & e^{\dis -\II\,k_i\,d_i}
 \end{array}\right)\,
 \left(\begin{array}{c}
    A^+_{i} \\ A^-_{i}
 \end{array}\right)
   = P_i\,
 \left(\begin{array}{c}
    A^+_{i} \\ A^-_{i}
 \end{array}\right).
\end{equation}
\begin{figure}
 \begin{center}
  \epsfbox{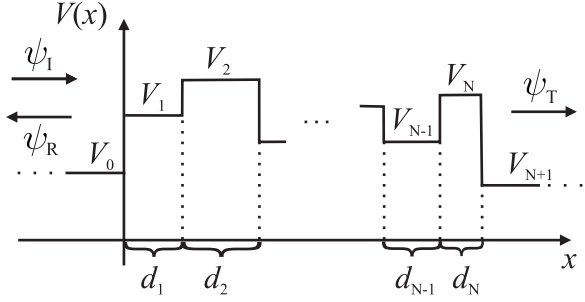}
 \end{center}
 \caption{\label{figuraPiecewisePotential}Piecewise constant potential with
  $N$ potential wells $V_i$ with $d_i$ as the corresponding width.
  $V_0$ and $V_{N+1}$ are the
  surrounding constant potential values extended to infinity.}
\end{figure}
Both the scattering and propagation matrices can be used to solve
the general problem of the scattering with a piecewise constant
potential with $N$ potential wells, as shown in
Figure~\ref{figuraPiecewisePotential}. The successive application
of the scattering and propagation matrices yield
\begin{equation}
 \left(\begin{array}{c}
    A^+_{0} \\ A^-_{0}
 \end{array}\right)
   = D_{0}^{-1}\,D_{1}\,
 \left(\begin{array}{c}
    A^+_{1} \\ A^-_{1}
 \end{array}\right)
   = D_{0}^{-1}\,D_{1}\,P_1\,D_{1}^{-1}\,D_{2}\,
 \left(\begin{array}{c}
    A^+_{2} \\ A^-_{2}
 \end{array}\right),
\end{equation}
and, in the most general form,
\begin{equation}
 \left(\begin{array}{c}
    A^+_{0} \\ A^-_{0}
 \end{array}\right)
   = M\,
 \left(\begin{array}{c}
    A^+_{N+1} \\ A^-_{N+1}
 \end{array}\right),
 \qquad
 M = D_{0}^{-1}\,
   \left(
    \prod_{i=1}^N
       D_{i}\,P_i\,D_{i}^{-1}
    \right)\,D_{N+1}.
\end{equation}
Both the reflection and transmission coefficients of the
scattering of a quantum particle, incoming from the left, with the
$N$-well potential is determined by the coefficients of the matrix
$M$,
\begin{equation}
 \left(\begin{array}{c}
    A^+_{0} \\ A^-_{0}
 \end{array}\right)
   = \left(\begin{array}{cc}
    M_{11} & M_{12} \\
    M_{21} & M_{22}
 \end{array}\right) \,
 \left(\begin{array}{c}
    A^+_{N+1} \\ 0
 \end{array}\right),
\end{equation}
where no backward moving particle can be found on the right side
of the potential, so $A^-_{N+1}=0$. The reflection and the
transmission coefficients~\cite{Liboff,Pedro} are given by
\begin{equation}
 R = \frac{\abs{A^-_{0}}^2}{\abs{A^+_{0}}^2} =
  \frac{\abs{M_{21}}^2}{\abs{M_{11}}^2},
 \qquad \mbox{and} \qquad
T = \frac{k_{N+1}\,\abs{A^+_{N+1}}^2}{k_{0}\,\abs{A^+_{0}}^2} =
  \frac{k_{N+1}}{k_{0}\,\abs{M_{11}}^2},
\end{equation}
respectively.

\begin{figure}
 \begin{center}
  \epsfbox{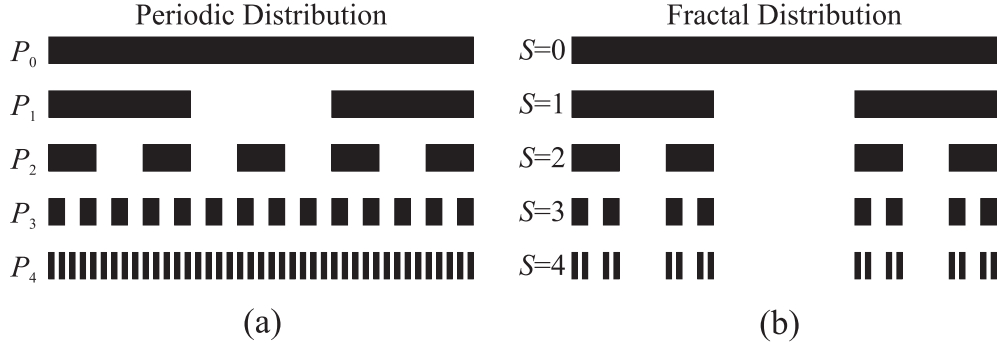}
 \end{center}
 \caption{\label{figuraPeriodicCantor}Finite periodic (a) and Cantor set quasiperiodic (b) potentials
  where the white and black regions denote the potential values $0$ and $\mathcal{V}$,
  respectively.}
\end{figure}

\section{Presentation of results}
\label{Results}

The simplest fractal potential is the Cantor set, shown in
Figure~\ref{figuraPeriodicCantor}(b), which can be obtained by
means of an iterative construction. The first step ($S=0$) is to
take a segment of unit length. The next one ($S=1$) is to divide
the segment in three equal parts of length $1/3$ and remove the
central one. In general, at the stage $S$, there are $2^S$
segments of length $3^{-S}$ with $2^S-1$ gaps in between. Stage
$S+1$ is obtained by dividing each of these segments into three
parts of length $3^{-S-1}$ and removing the central ones. In
Figure~\ref{figuraPeriodicCantor}(b), only the four first stages
are shown for clarity. Note that the $S$-th stage Cantor set
pre-fractal can be interpreted as a quasiperiodic distribution of
segments which can be obtained by removing some segments in a
finite periodic distribution as shown in
Figure~\ref{figuraPeriodicCantor}(a). This distribution at stage
$p_M$ has $(3^M-1)/2+1$ potential barriers of length $3^{-M}$,
separated by potential wells of the same length, so the ``period"
of this finite structure is $\Lambda=2\,\cdot\,3^{-M}$.

The scattering problem for both the quasiperiodic, Cantor set,
pre-fractal potential, and the finite periodic potential can be
easily solved by means of the matrix transfer theory presented in
Sec. 2. It is standard to normalize both the energy and the height
of the potential barrier by the period $\Lambda$, introducing the
non-dimensional variables
\[
 \phi=\Lambda\,\frac{\sqrt{2\,m\,E}}{\hbar}, \qquad \mbox{and} \qquad
 \phi_\mathcal{V}=\Lambda\,\frac{\sqrt{2\,m\,\mathcal{V}}}{\hbar}.
\]

\begin{figure}
 \begin{center}
  \epsfbox{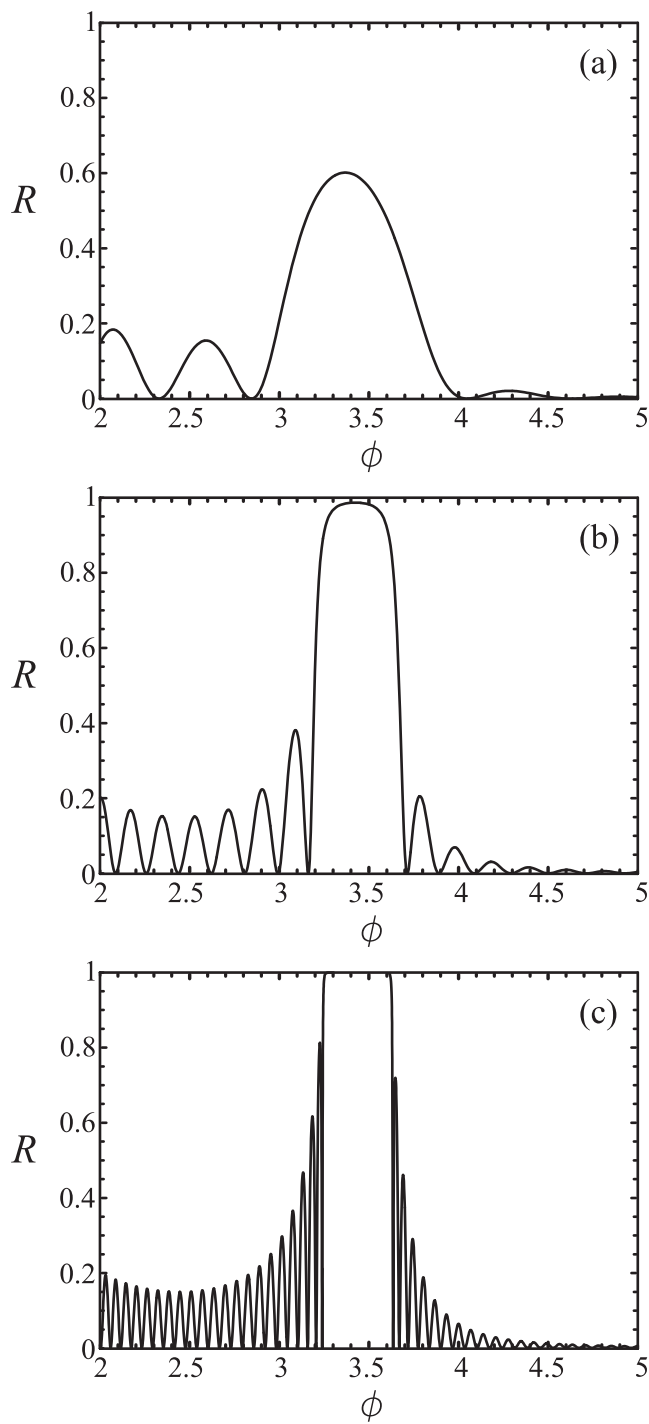}
 \end{center}
 \caption{\label{figuraRforPeriodic}Scattering reflection coefficient for the
  finite periodic potentials of stages $p_2$ (a), $p_3$ (b),
  and $p_4$ (c) as a function of the normalized energy $\phi$
  for the potential $\phi_\mathcal{V}=2$.}
\end{figure}

\begin{figure}
 \begin{center}
  \epsfbox{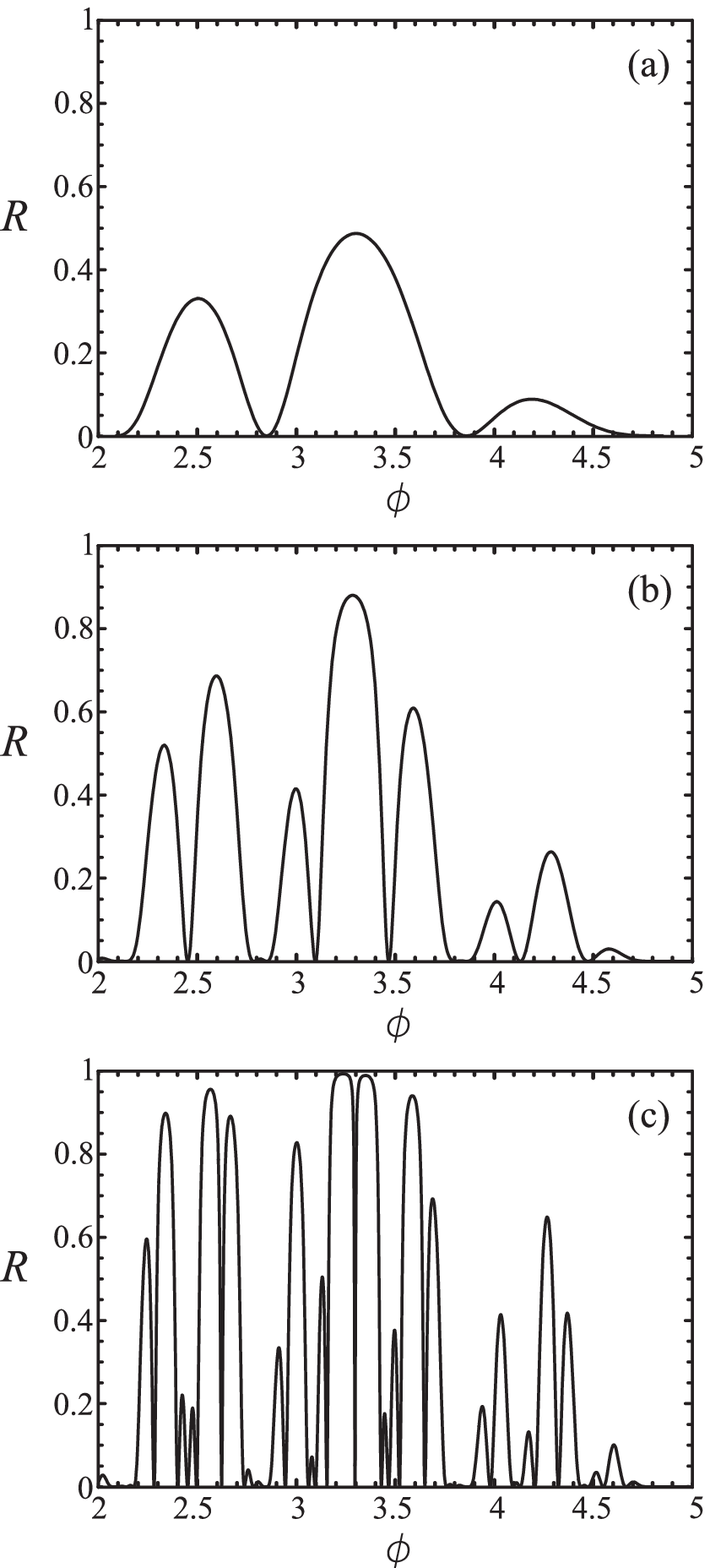}
 \end{center}
 \caption{\label{figuraRforCantor}Scattering reflection coefficient for the
  Cantor set pre-fractal potentials of stages $S=2$ (a), $S=3$ (b),
  and $S=4$ (c) as a function of the normalized energy $\phi$
  for the potential $\phi_\mathcal{V}=2$.}
\end{figure}

Figures~\ref{figuraRforPeriodic} and~\ref{figuraRforCantor} show
the reflection coefficient, $R$, for the finite periodic potential
and Cantor set fractal potential, respectively, around the
interval which contains the first band gap of the infinite
periodic one. Using the standard Kronig-Penney
model~\cite{KronigPenney2}, this band gap can be numerically
calculated yielding $3.2519 < \phi < 3.6222$ for the potential
$\phi_\mathcal{V}=2$.
In this energy interval, a Bloch wavefunction does not propagate
in a infinite periodic potential and, therefore, the transmission
coefficient should vanish ($R=1$). Only evanescent wavefunctions
characterized by a complex wavevector, $k$, are solutions of the
Schr\"odinger equation. For this reason, when the number of
periods is finite, the quantum particle may pass through the
potential distribution by the tunnelling effect.
Figure~\ref{figuraRforPeriodic} shows that the reflection
coefficient approaches unity as the number of periods in the
spatial interval increases, illustrating the process of appearance
of the band gap of the (full) periodic structure. Although, at the
graphical resolution of Figure~\ref{figuraRforPeriodic}(c) the
value $R=1$ is apparently reached, the reflection coeficient is
always smoller than unity in finity peridodic structures.

Figure~\ref{figuraRforCantor} shows the reflection coefficient for
the Cantor set pre-fractal potential for $S=2$ (top), $S=3$
(middle) and $S=4$ (bottom). It is shown that the reflection at
each higher stage is a modulated version of that associated with
the previous stage. That is, the reflection spectrum exhibits a
characteristic fractal profile that reproduces the self-similarity
of the potential distribution. In fact, any wide peak at stage $S$
is transformed into three narrower and taller peaks at stage
$S+1$. Zero reflection from these fractal quantum potential occurs
at specific discrete energies, while near total reflection is
possible at other discrete energies. Comparing
Figures~\ref{figuraRforPeriodic} and~\ref{figuraRforCantor}, an
increasing number of zeros inside the band gap is observed. These
zeros represent resonances due to the presence of ``defects" in
the quasiperiodic potential obtained by removing some segments in
the finite periodic sequence.

\section{Conclusions}
\label{Conclusions}

The transfer matrix method is becoming the standard method for the
calculation of the tunnelling of quantum particles on constant
piecewise potentials because it can be used for simple,
textbook-like problems and as a numerical method for computer
simulations. This procedure has been applied to Cantor set fractal
potentials, which are constant value potentials with support on a
Cantor set. For pre-fractals, the $S$-th stage fractal, the
reflection coefficient was numerically calculated and compared
with that of a finite periodic potential of the same period. The
appearance of the first band gap of the Kronig-Penney model in the
finite periodic potential has been illustrated. The reflection
coefficient for the Cantor set potential is self-similar.

The transfer matrix method presented in this paper can be easily
adopted in computer laboratories for undergraduate quantum
mechanics courses, providing a powerful method for developing
students skill on physics by means of computational tools.
Furthermore, fractal geometry is a highly motivating topic for the
students providing a great opportunity to undertake projects
closely related to research ones.

\ack{The authors are thankful to Prof. Sarira Sahu from the
Instituto de Ciencias Nucleares at the Universidad Aut\'onoma de
M\'exico, M\'exico, and Prof. Juan I. Ramos from the Universidad
de M\'alaga, Spain, for their valuable comments and suggestions.
J.A. Monsoriu and P. Fern\'andez de C\'ordoba were supported by
the Plan Nacional I+D+I under project TIC 2002-04527-C02-02
(Spain). F.R. Villatoro was supported by Project BFM2001-1902 from
the Direcci\'on General de Investigaci\'on, Ministerio de Ciencia
y Tecnolog\'{\i}a, Spain. Part of this work was done during the
visit of J.A. Monsoriu to the Universidad de M\'alaga with a grant
from the Universidad Polit\'ecnica de Valencia, under the
``Programa de Incentivo a la Investigaci\'on de la UPV 2004".}

\newcommand{\bookref}[4]{#1 #2 {\em {#3}} (#4)}

\newcommand{\paperref}[5]{#1 #2 {\em {#3}} {\bf #4} #5}

\end{document}